# Decorated dislocations against phonon propagation for thermal management

Stefanos Giaremis[1], Joseph Kioseoglou[1], Mykola Isaiev[2], Imad Belabbas[3], Philomela Komninou[1] and Konstantinos Termentzidis[4]

[1] Department of Physics, Aristotle University of Thessaloniki, GR-54124 Thessaloniki, Greece

[2] Université de Lorraine, LEMTA, CNRS-UMR7563, BP 70239, 54506 Vandœuvre Cedex, France

[3] Equipe de Cristallographie et de Simulation des Matériaux, Laboratoire de Physico-Chimie des Matériaux et Catalyse, Faculté des Sciences Exactes, Université de Bejaia 06000, Algeria

[4] Univ. Lyon, CNRS, INSA-Lyon, Université Claude Bernard Lyon 1, CETHIL UMR5008, Villeurbanne, France

**Abstract:** The impact of decorated dislocations on the effective thermal conductivity of GaN is investigated by means of equilibrium molecular dynamics simulations via the Green-Kubo approach. The formation of "nanowires" by a few atoms of In in the core of dislocations in wurtzite GaN is found to affect the thermal properties of the material, as it leads to a significant decrease of the thermal conductivity, along with an enhancement of its anisotropic character. The thermal conductivity of In-decorated dislocations is compared to the ones of pristine GaN, InN, and a random and an ordered $In_xGa_{1-x}N$ alloy, to examine the impact of doping. Results are explained by the stress maps, the bonding properties and the phonon density of states of the aforementioned systems. The decorated dislocations engineering is a novel way to tune, among other transport properties, the effective thermal conductivity of materials at the nanoscale, which can lead to the manufacturing of interesting candidates for thermoelectric or anisotropic thermal dissipation devices.

**Introduction**

GaN has been established as a material of choice among III-V semiconductors for light emitting diodes (LEDs), laser diodes (LD), high – power high – frequency transistors, photo-voltaic and piezoelectric devices [1, 2, 3]. The modification of its emission and absorption properties can be achieved by alloying with In [4, 5, 6], with significant recent advances reported also in the field of thermoelectric (TE) applications [7]. However, in both GaN and $In_xGa_{1-x}N$ c-plane heteroepitaxially grown layers, a high number of threading dislocations (TDs) has been observed [8, 9, 10, 11]. The presence of TDs in these nitride – based materials crucially affect their electronic properties [12, 13] and thus device efficiency [14, 15] as they act as nonradiative carrier recombination centers [16], while high density of dislocations relates with reduced lifetimes in optoelectronic devices [17] and more recently, reduction of thermal conductivity in established TE nanostructured materials [18]. The latter effect will be discussed in detail in the following sections.

Dislocations in general are known to have an impact on the thermal properties of a material, as they induce thermal resistivity due to elastic phonon scattering [19, 20, 21, 22]. In particular, the very high density of dislocations observed in III-nitride based devices [8] is found to strongly influence their performance and thermal transport behavior [23, 24, 25, 26], inducing high thermal transport anisotropy [27]. In GaN, the behavior of thermal conductivity, k, is characterized by two regimes, depending on the density of dislocations [24, 25, 26, 19, 28]; thermal conductivity is independent of the density of dislocations in the low-density regime, while it presents a logarithmic dependence on the high density regime. The threshold for the density of dislocations between these two regimes is reported to be at $10^7$ cm$^{-2}$, based on experimental measurements [24] and $10^{10}$ - $10^{11}$ cm$^{-2}$ from computational studies [25, 26, 28]. Furthermore, the presence of dislocations in wurtzite bulk GaN is associated to a suppression of the TA modes, resulting thus to a reduction of k [28]. It should be noted that, although the majority of the threading dislocations in wurtzite GaN are of the edge type, the most crucial dislocations either for the optoelectronic or the thermal properties of the devices are screw ones [19, 29], as they are found to yield a twofold decrease of k compared to their edge counterparts [19].

Although dislocations are regarded mainly as undesired side effects in nanostructured semiconductors, their manipulation can lead to interesting morphological exploits for the modification of materials properties at the nanoscale. As dislocations are one-dimensional structures similar to nanowires (NWs), they introduce a transversal translational discontinuity, which implies a long-range strain field attracting impurities and consequently, modifying the stoichiometry in their vicinity and thus the overall electronic structure and the thermal behavior of the material. This so-called "dislocation technology" scheme has been established almost ten years ago, and it is based on the decoration of dislocation cores by specific species acting as impurities and results in formation of unidirectional functional NWs inside bulk materials [30]. "Dislocation technology" is highly promising, especially in materials with high dislocation density and it is expected to attribute distinct properties in commonly used materials [31, 32]. In III-N semiconductors the density of TDs is quite high, and the construction of conductive NWs by concentrating metallic atoms along the threading dislocations of GaN [33] and the almost insulator AlN [34, 35, 36] has been already verified. In $In_xGa_{1-x}N$ alloys, as In atoms are larger than Ga ones, they are found to segregate to the tensile region of the strain field of dislocations [37]. In particular, it has been proven that the energetically most favorable configuration for In atoms, assuming they are mobile enough during growth, is to bind to cores of c-type screw dislocations, forming an In–rich region rather than the equivalent $In_xGa_{1-x}N$ core [38, 39, 40]. The presence of In atoms in the core of c-screw dislocations in GaN is also found to further stabilize the core configuration [40]. However, there is a noticeable lack of studies addressing the thermal properties of such nanostructures in semiconductors and the influence of the decoration on k in particular.

The understanding and manipulation of the above effects is important for the optimization and development of numerous novel III-nitride based applications. As the characteristic length scale of nanostructured devices is continuously decreasing, detailed understanding and manipulation of the heat flow mechanisms at the nanoscale is crucial [41]. Furthermore, structural asymmetries and nonlinear lattice structures are desired for thermal diodes and transistors [42, 43], while anisotropic heat flux can be exploited for devices such as thermal cloaks [44]. Another interesting field of applications is thermoelectricity, a phenomenon which enables the conversion of heat into electric energy via the Seebeck effect. The development of TE materials is of great importance due to their use in renewable energy applications and waste heat recovery in home appliances, industry, transport and extreme environments, such as space [45, 46]. An efficient TE material exhibits high Seebeck coefficient, high electrical conductivity and low thermal conductivity. One approach for increasing the figure of merit (zT), a measurement of thermoelectric performance of the material, is by minimizing k. There are numerous works in the literature towards this direction, with the more recent ones involving the reduction of thermal conductivity by phonon scattering at dense dislocations [47, 48, 49] and particularly of the screw type [18]. High performance TE materials currently include $Bi_2Te_3$ based superstructures, SiGe, skudderudites, half-Heusler, Zintl phase and clathrate alloys [50]. III-nitrides are promising candidates as high temperature TE materials, as they exhibit properties such as wide bandgap, mechanical robustness and chemical stability at high temperatures [51]. In particular, the use of GaN for this purpose has been proven viable [52, 53] and it was demonstrated in experimental applications such the integration of n-type thin film GaN based thermoelectric micro power generator device suitable for on-chip integration [54]. Moreover, Pantha et al [55] has reported a zT of 0.23 at 450K for InGaN films at 36%

In composition, which is comparable to the performance of SiGe alloys at the same temperature. Sztein *et al* [56] reported values of zT=0.04 at room temperature with a maximum value of 0.34 at 875K for n-type InGaN films at 17% In composition, while Kucukgok *et al* [57] reported a zT of 0.072 at room temperature at 20% In content. Similar results were also recently reported for InGaN/GaN heterostructured films by Surender et al [58]. These later works demonstrated the improved TE performance of InGaN compared to GaN, as the alloying with In was found to reduce k due to mass disorder and local strain field scattering without negative impact on the electrical conductivity and Seebeck coefficient. Phonon scattering by nanometer-scale compositional inhomogeneities was also demonstrated as a mechanism for the reduction of k in InGaN alloys for Ga-rich and In-rich compositions [59]. More recently, a record value of zT=0.86 at room temperature was achieved for $In_{0.32}Ga_{0.68}N$ on ZnO substrate by Feng *et al* [7], due to the further enhancement of electrical conductivity by oxygen co-doping besides the reduction of k due to alloy scattering.

Regarding the field of theoretical predictions of thermal transport properties in the nanoscale, descriptions of Dislocation – Phonon Interactions (DPI) were firstly attempted five decades ago by by Klemens and Carruthers with the static strain field DPI models based on nonlinear elasticity theory [60, 61]. However, these models were proved to be problematic, as they deviated from universally experimental data [62, 63]. An explanation for this discrepancy was that the validity of the perturbations analysis and the Born approximation due to the long range nature of the dislocation strain field [63]. Dynamic scattering DPI models describe a mechanism also referred to as "fluttering", which was a later advancement in the static DPI theory and according to which, a dislocation absorbs an incident phonon, vibrates and subsequently re-emits a phonon [62, 64, 65]. Recent advances towards a unified description of DPI are based on phonon renormalization, with a model that treats the DPI in the framework of quantum field theory and dislocations as quantized fields, named "dislons" [63, 66]. *Ab initio* methods have been proven successful for the prediction of thermal transport [23, 67, 68, 69, 70], although very few of them address the effect of dislocations [23, 68], due to the large volume of the supercells required for the modeling of these defects. Molecular Dynamics (MD) simulations, on the other hand, are an established technique for this task due to their applicability for upscaling [71, 72], which has already yielded successful results for bulk and nanostructured materials in the family of III-nitrides [19, 73, 74, 75].

The present study aims to give insights on the impact of the formation of In–rich screw dislocation cores on the thermal properties of InGaN alloys at the high dislocation density regime, by using classical Equilibrium Molecular Dynamics (EMD) atomistic simulations. First, the computational procedure that was implemented and the case studies under consideration will be presented in detail, followed by an analysis of the stress fields, the dependence of k on the dislocations density and the phonon density of states for each system. These re-sults aim towards a better understanding of the combined effect of phonon scattering in dislocations and mass disorder in wurtzite GaN and InGaN alloy to the effective thermal properties of these materials, leading potentially to the optimization and the design of novel InGaN based devices.

**Computational Details**

Calculations were performed on parallelepiped volumes containing approximately 19200 atoms each. Four screw TDs with single 6 atom ring core configurations [76, 77, 78] with their lines along the [0001] direction and alternating signs were constructed. This manner was chosen as it facilitates the mutual annihilation between dislocation pairs and prevents core destabilization while applying periodic boundary conditions on the simulation box due to the asymmetric morphology of the strain field associated with this particular type of dislocations. The modelling of the dislocations was achieved by the imposition of a displacement field of screw dislocations, as described by the theory of linear elasticity [79]:

$$u_z = \frac{b}{2\pi} \arctan \frac{y}{x} \qquad (1)$$

expressed in a Cartesian orthonormal coordinate system with z lying along the dislocation line, on the perfect wurtzite GaN crystal. *b* stands for the magnitude of the Burgers vector where *b*=*c*=5.28Å. The origin of the displacement field was chosen according to the methodology implemented by Termentzidis et al [19] and corresponds to a dislocations' density of $10^{11}$cm$^{-2}$, close to the computational threshold between the high and the low density regime. For the aforementioned structure, three separated cases were investigated, where Ga atoms were substituted by In ones on the center of the dislocation core and on its first and second neighbors, respectively. These cases correspond to total In concentration levels of 2.5%, 5% and 10% on the supercells, or expressed via the stoichiometric formula, $In_xGa_{1-x}N$, with x=0.025, 0.05 and 0.1, respectively.

To isolate the impact of the In–decorated screw dislocations on k, further calculations were performed in pristine bulk InN, GaN and $In_xGa_{1-x}N$ alloy supercells, modelled as parallelepiped volumes containing approximately 11000 atoms. Regarding the $In_xGa_{1-x}N$ alloy, two cases where considered. The first one involves calculations on pristine, bulk systems with the same In concentration, x, with the configurations containing the decorated dislocations. In this case, the desired In concentration on each subcase was achieved by replacing Ga atoms with In ones in a random manner, thus resulting to a random alloy. On the second case, calculations where performed with higher In concentration levels (x=0.0625, 0.125, 0.1875 and 0.25). In these subcases, the desired In compositions where achieved by the replacement of Ga atoms by In ones only on the energetically preferable lattice sites, based on the work of Pavloudis et al. [80], thus resulting to a kind of ordered alloy. In all of the above cases, lattice constants of pristine wurtzite GaN were rescaled for the accommodation of the In-induced stress.

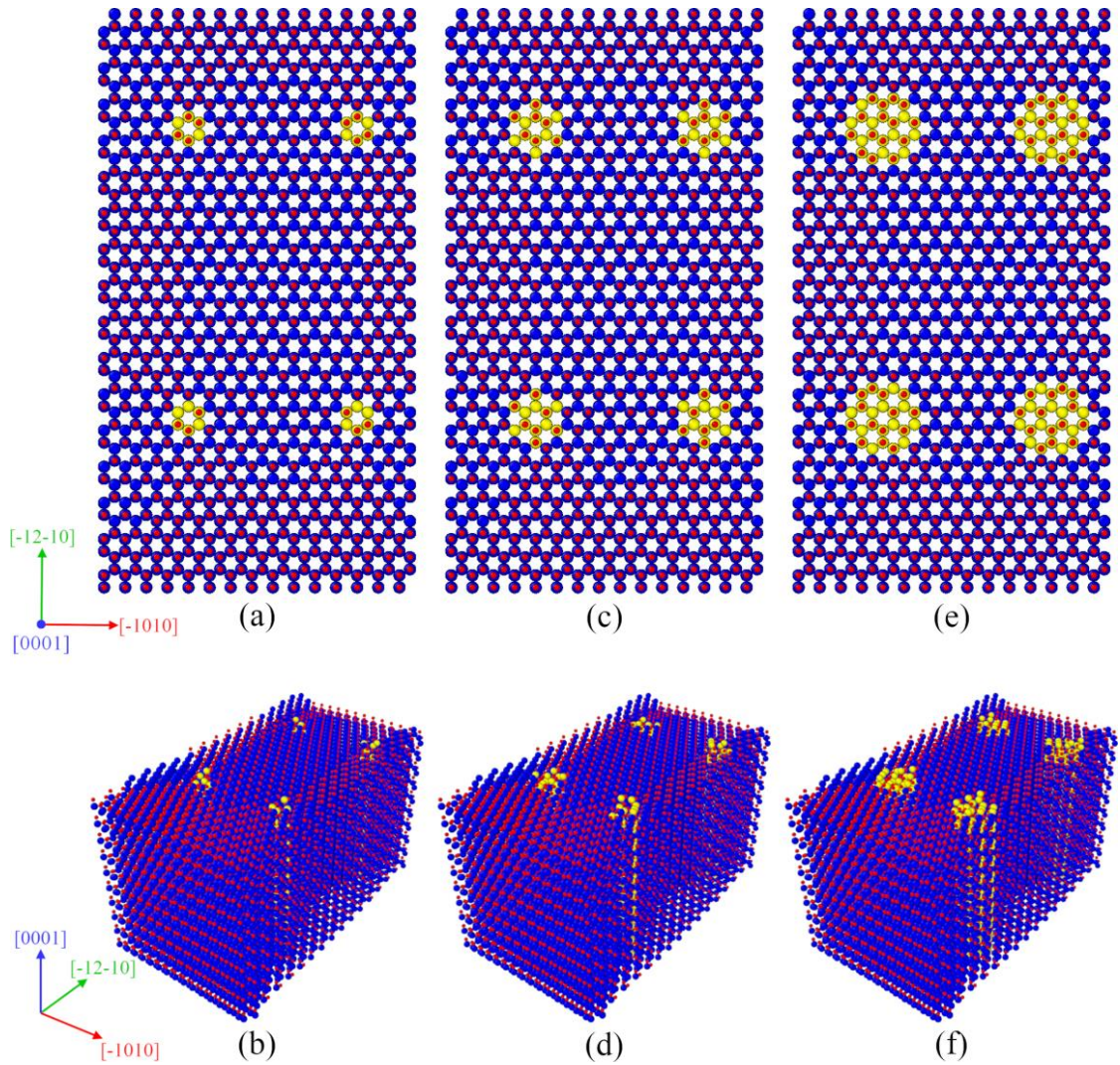

**Figure 1: Supercells containing four In-decorated screw dislocations each, with a total In content of x=0.025 (a), (b), 0.05 (c), (d) and 0.1 (e), (f). View along <0001> (a), (c), (e) and from a 3D perspective (b), (d), (f). In atoms are represented as yellow, Ga atoms as blue and N atoms as red.**

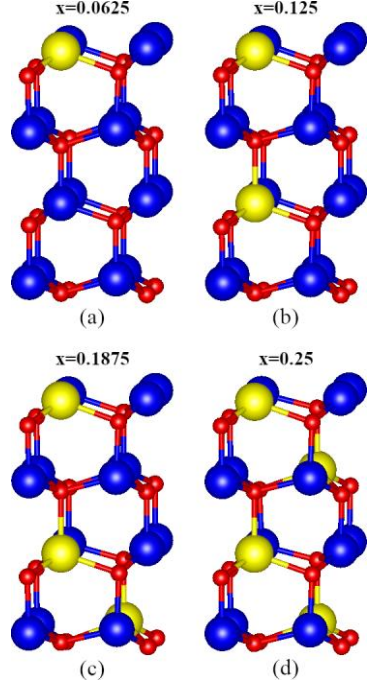

**Figure 2:** Supercells of 16 atoms used by Pavloudis et al. [80] for the investigation of the energetically preferable lattice sites for the replacement of Ga atoms by In, resulting to an ordered $In_xGa_{1-x}N$ (x=0.0625, 0.125, 0.1875 and 0.25) alloy. The energetically favorable configuration was found for the case of x=0.25. In atoms are represented as yellow, Ga atoms as blue and N atoms as red.

In molecular dynamics simulations of III-N dislocations, a variety of interatomic potentials have been used, e.g. [81, 82, 83]. However, the most popular ones are the Tersoff [84] and Stillinger-Weber (SW) [85] based potentials. However, the SW potential owing to its simple mathematical formulation has been established as the most suitable for computationally and timely expensive large-scale calculations [86]. For instance, on a recent study, the thermal properties of the edge and screw dislocations in GaN have been elucidated by using the SW interatomic potential [19].

For the calculation of the thermal conductivity, EMD simulations were performed by using LAMMPS [87], with a modified Stillinger Weber interatomic potential [88] and the Green-Kubo approach. In the latter approach, heat flux fluctuations in equilibrium are connected to thermal conductivity via the fluctuation-dissipation theorem:

$$\kappa_z = \frac{V}{k_B T^2} \int_0^\infty \left\langle \vec{J}(t) \cdot \vec{J}(0) \right\rangle dt \qquad (2)$$

where $V$ is the volume of the system, $T$ is the equilibrium temperature and $k_B$ is the Boltzmann constant. $\vec{J}$ is the equilibrium heat flux vector, calculated via:

$$\vec{J} = \frac{1}{V}\left[ \sum_i E_i \vec{v}_i + \frac{1}{2}\sum_{i<j}\left( \vec{F}_{ij} \cdot \left(\vec{v}_i + \vec{v}_j\right)\right) \right] \qquad (3)$$

where $E_i = \frac{1}{2}\left(m|\vec{v_i}|^2 + \sum_i \varphi_{ij}\right)$ is the energy of each atom, $\varphi_{ij}$ is the interatomic potential, $\vec{F_{ij}}$ is the interatomic force vector between atoms $i$ and $j$ and $\vec{v_i}$ is the velocity vector of atom i (see also Khadem et al [72]). The time step used for the calculation of the correlations in the microscopic heat flux is 2 ps, while the correlation data for the preceding samples are computed for 40 ps. The timestep of the total simulation was set to 0.5 fs. The full computational process for the EMD calculation involved thermalisation of the system as an NVT ensemble at 300 K for 200 ps, equilibration as NVE for 2 ns and then averaging for 4–5 ns to ensure the convergence of the integral part of Eq. (3), known as the Heat Auto-Correlation Function (HCAF). Simulations were performed using 10 different seeds of initial velocity for the simulated particles to reduce statistical error.

Stress fields were also calculated by LAMMPS via the following formula [89]:

$$S_{ab} = \left[ -mu_a u_b + \frac{1}{2}\sum_{n=1}^{N_p}\left(r_{1,a}F_{1,b} + r_{2,a}F_{2,b}\right) + \frac{1}{2}\sum_{n=1}^{N_b}\left(r_{1,a}F_{1,b}\right) + \right.$$
$$\frac{1}{3}\sum_{n=1}^{N_a}\left(r_{1,a}F_{1,b} + r_{2,a}F_{2,b} + r_{3,a}F_{3,b}\right) +$$
$$\frac{1}{4}\sum_{n=1}^{N_d}\left(r_{1,a}F_{1,b} + r_{2,a}F_{2,b} + r_{3,a}F_{3,b} + r_{4,a}F_{4,b}\right) +$$
$$\left. \frac{1}{4}\sum_{n=1}^{N_i}\left(r_{1,a}F_{1,b} + r_{2,a}F_{2,b} + r_{3,a}F_{3,b} + r_{4,a}F_{4,b}\right) \right]$$

(4)

Eq. (4) yields the value of the stress tensor for a given atom in units of pressure times volume. For the estimation of the total pressure per atom, each value was divided by the volume of the primitive cell. Indices $a,b$ take values $x,y,z$, as $S_{ab}$ is a symmetric tensor with 6 components. The first term of Eq. (4) is the kinetic energy contribution for the given atom, where $m$ is the mass and $u$ the velocity of each atom. This term is equal to zero in the present case. The second term is a pairwise energy contribution, were $N_p$ is the number of neighbors for the given atom, $r_{1,2}$ are the positions and $F_{1,2}$ the forces applied on each atom of the pair. Third, fourth and fifth terms arise from angular ($N_a$), dihedral ($N_d$) and improper interactions ($N_i$) that the given atom is part of.

**Results and Discussion**

The effective k of In-decorated dislocations compared to the random $In_xGa_{1-x}N$ alloy with the same In content, as a function of In content, is depicted in Figure 3. Furthermore, Table I presents the results for the same system sizes of pristine bulk InN and GaN with the same simulation setup to help the discussion. Bulk GaN exhibits a significantly high thermal conductivity compared to bulk InN, fact that can be associated to the length and the strength of the Ga-N bond, which is shorter and stronger compared to the In-N one (Table II). Furthermore, In is a heavier element than Ga, and in general compounds comprised of light elements are linked to high values of k [90]. The reduction of k due to the presence of In–decorated screw dislocations compared to the values for pristine bulk GaN and InN is substantial. Thermal conductivity of random $In_xGa_{1-x}N$ alloys for the same In concentration, x, with the cases of In-decorated screw dislocations in GaN was found to be even lower. Results for the $In_xGa_{1-x}N$ alloy are in good agreement with other theoretical predictions [91] and experimental data at 300K [56]. For smaller In concentration levels, the formation of In-decorated dislocations leads to a twofold increase of k compared to the respective values for the random alloy. This effect seems also to be more pronounced for lower In concentration levels. However, for In composition close to x=0.1, In-decorated dislocations are hindering thermal conductivity further than the pristine bulk random alloy with the same In composition.

**Table I: Values of k for pristine InN and GaN.**

| k component ($Wm^{-1}k^{-1}$) | InN | GaN |
|---|---|---|
| **in - plane** | 55.25 | 142 |
| **c - axis** | 58 | 160 |
| **effective** | 56.625 | 151 |

**Table II: Crystallographic parameters for InN and GaN as calculated with the modified SW potential [88].**

|  | Modified SW potential [88] | | Experimental / *ab initio* | |
|---|---|---|---|---|
|  | InN | GaN | InN | GaN |
| **Bond length (Å)** | 2.254 | 1.954 | 2.156 [92] | 2.06 [93] |
| ***a*(Å)** | 3.681 | 3.190 | 3.538 [94] | 3.189 [95] |
| ***c/a*** | 1.633 | 1.633 | 1.612 [94] | 1.626 [95] |
| **Binding energy (eV)** | −7.976 | -8.682 | - | - |

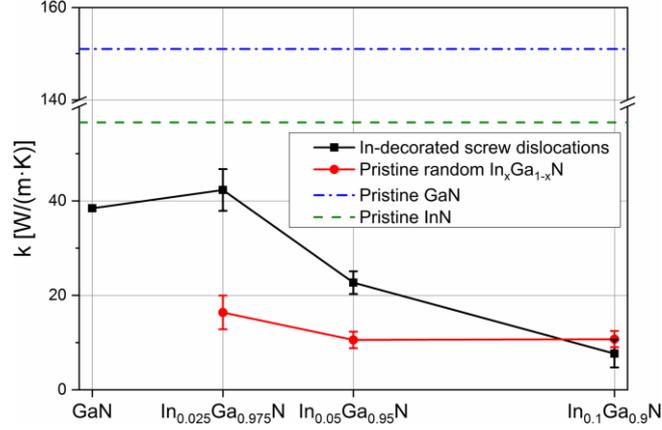

**Figure 3:** Effective k for the case of In-decorated screw dislocations in bulk GaN (black) for x=0, 0.025, 0.05 and 0.1, and pristine $In_xGa_{1-x}N$ random alloy (red) for x=0.025, 0.05 and 0.1, along with the values of effective k for pristine GaN (blue dash dotted) and InN (green dashed) for comparison.

In Figure 4a, results of directional k calculations are presented for the first case which corresponds to the In-decorated screw dislocations. In–plane refers to (0001) and c axis to <0001>. The anisotropic character of k is evident and more pronounced for In concentration levels of up to approximately x=0.05, while tending to be eliminated for higher In concentration levels. In Figure 4b, the directional behavior of k is presented for the case of random $In_xGa_{1-x}N$ alloys. k appears to be less anisotropic compared to the In-decorated screw dislocations on bulk GaN, as shown on the previous case, with the anisotropy tending to be completely eliminated for In concentration levels close to x=0.05. In Figure 4c, the behavior of the k is demonstrated for the case of ordered $In_xGa_{1-x}N$, modeled via the substitution of Ga atoms with In ones in the energetically preferable sites on pristine bulk GaN, for In concentration levels up to x=0.1875. k appears to be relatively isotropic, too, compared to the case of In-decorated screw dislocations on bulk GaN, with a monotonous decrease for In concentration levels up to x=0.1875. By combining these values with the results presented in Figures 3 and 4a, b and c, the impact of In-decorated screw dislocations and In doping on pristine GaN on thermal conductivity is highlighted, as they appear to lead to a significant reduction of both in-plane and c axis components of thermal conductivity in all cases, compared to the binary compounds. This result is consistent with recent similar studies, showing evidence that ordering is associated with higher values of thermal conductivity in alloys [96].

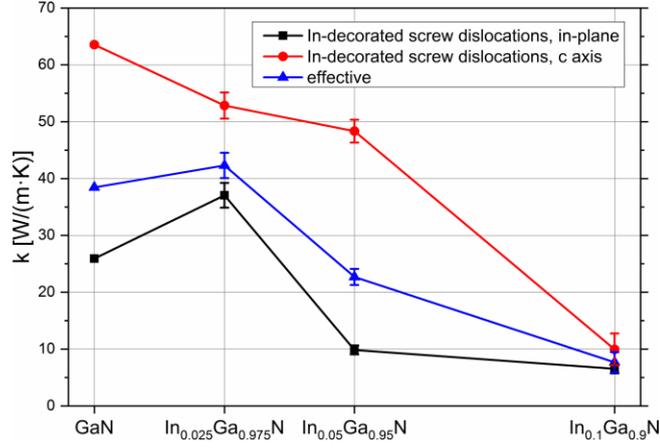

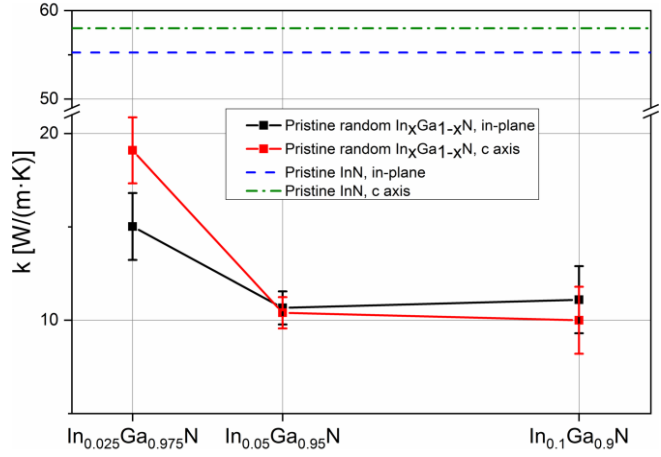

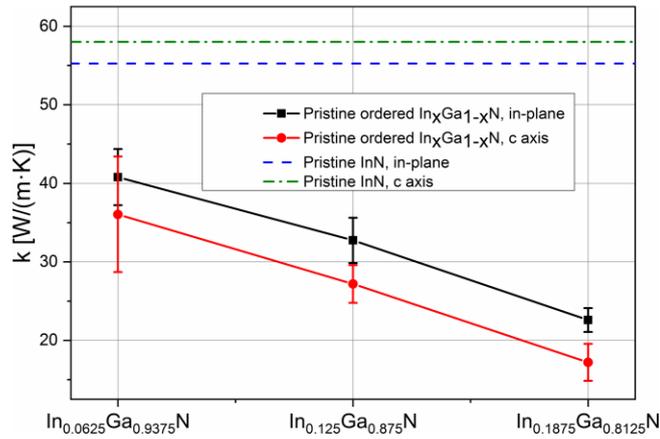

**Figure 4: (a):** In–plane (black) and c axis (red) components of k along with the respective values of effective k (blue), for the case of In-decorated screw dislocations in bulk GaN for x=0.025, 0.05 and 0.1. **(b):** In–plane (black) and c axis (red) components of the k of pristine $In_xGa_{1-x}N$ random alloy for x=0.025, 0.05 and 0.1, along with the values of in-plane (blue dashed) and c axis (green dash dotted) components of k in pristine bulk InN for reference. The respective values for pristine GaN are 142 $Wm^{-1}k^{-1}$ and 160 $Wm^{-1}k^{-1}$. **(c):** In – plane (black) and c-axis (red) components of k of ordered pristine $In_xGa_{1-x}N$ alloy, for x up to 0.1875, along with the values of in-plane (blue dashed) and c axis (green dash dotted) components of k in pristine bulk InN for reference. The respective values for pristine bulk GaN are 142 and 160 $Wm^{-1}k^{-1}$. The desired In concentration in the pristine ordered compounds was achieved by replacing Ga atoms with In ones in bulk wurtzite GaN on the energetically favorable lattice sites [80].

From Figure 5 it is evident, due to the morphology of the stress field associated to the In-decorated screw dislocations, as calculated via Eq. (4) that In content higher than x=0.05, fully segregated on the dislocation core leads to high stress on the bulk which cannot be properly accommodated, effect that is being furthermore amplified by the interaction between the opposite sign dislocations. The later effect is inevitable at the given dislocation density and was still observed after strict stress relaxation of the supercells. The $S_{zz}$ component can be attributed to both the influence of the screw dislocation and the In atoms, while the in-plane components ($S_{xx}$, $S_{yy}$) are only due to mass disorder from the presence of In atoms, since screw dislocations do not affect in-plane components [79]. This observed lack of stress accommodation can be explained by the greater length of the In-N bond compared to the Ga-N. It is also consistent with previous studies, reporting a partial release of the bond length deformation, but a significant bond angle deformation for In atoms close to the dislocation cores [40]. This effect can be also linked to the decreasing anisotropy between in-plane and c axis components of thermal conductivity for the same system, with increasing In content greater than x=0.05, as presented on Figure 4a. Tensile strain has been previously linked to a continuous decrease of k in silicon and diamond nanostructures [97], therefore the increase of the volume of regions under tensile stress with increasing of In content can be associated with the decrease of effective k in Figure 4a.

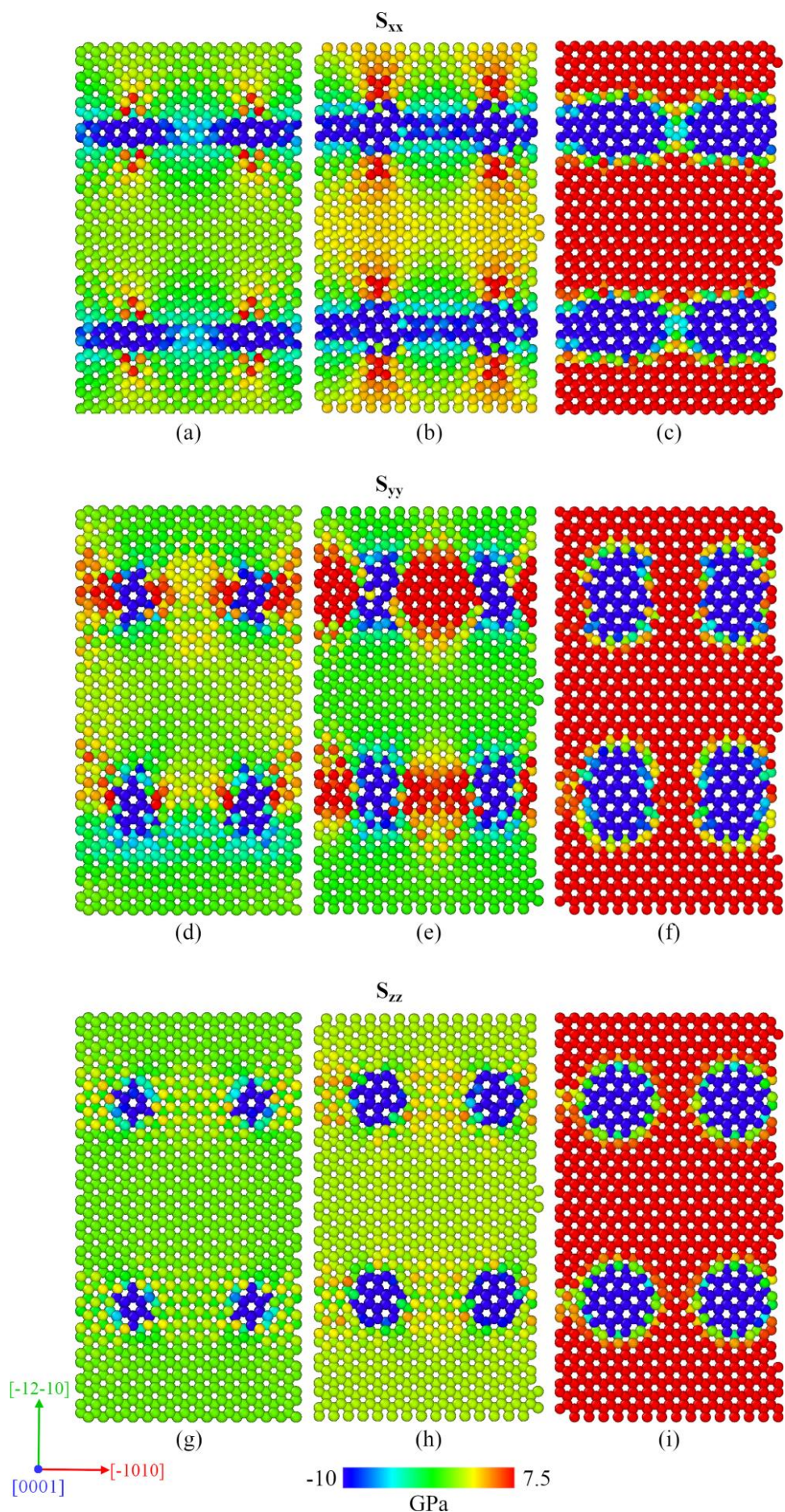

**Figure 5: xx (a), (b), (c), yy (d), (e), (f) and zz (g), (h), (i) components of the stress tensor for conductivity for the case of In-decorated screw dislocations in bulk GaN for x=0.025 (a), (d), (g), 0.05 (b), (e), (h) and 0.1 (c), (f), (i). x component corresponds to [-1010], y to [-12-10] and z to [0001] crystallographic directions, respectively.**

In Figure 6, the calculated values of k for the cases of In-decorated dislocations in bulk GaN, pristine random $In_xGa_{1-x}N$ and pristine ordered $In_xGa_{1-x}N$ are plotted as a function of the normalized density, $\rho_{normalized}$, defined as:

$$\rho_{normalized} = \frac{\rho}{x\rho_{InN} + (1-x)\rho_{GaN}} \quad (5)$$

where ρ is the respective density of each structure, x is the In content and $\rho_{InN}$, $\rho_{GaN}$, are the densities of InN and GaN, respectively. The density of each structure is calculated by the total mass divided by the volume of each supercell. k appears to have a monotonous dependence on $\rho_{normalized}$ in all three cases. For the cases of In-decorated screw dislocations in bulk GaN and pristine $In_xGa_{1-x}N$ random alloys, k is increasing monotonously with increasing $\rho_{normalized}$, with a steeper increase for the case of In-decorated dislocations. On the other hand, the inverse effect is observed for the pristine ordered $In_xGa_{1-x}N$ alloys, as k is decreasing monotonously with increasing $\rho_{normalized}$. This change in the form of the dependence between the three cases demonstrates the combined impact of ordering and mass on k. The steeper increase of k in Figure 6a compared to Figure 6b can be attributed to the stronger impact of mass in defected structures, while ordering seems to lead to an inverse effect, with k being decreased linearly with increasing $\rho_{normalized}$. These results are consistent with previous studies, as the dependence of k on mass is reported on similar systems [90, 98, 99].

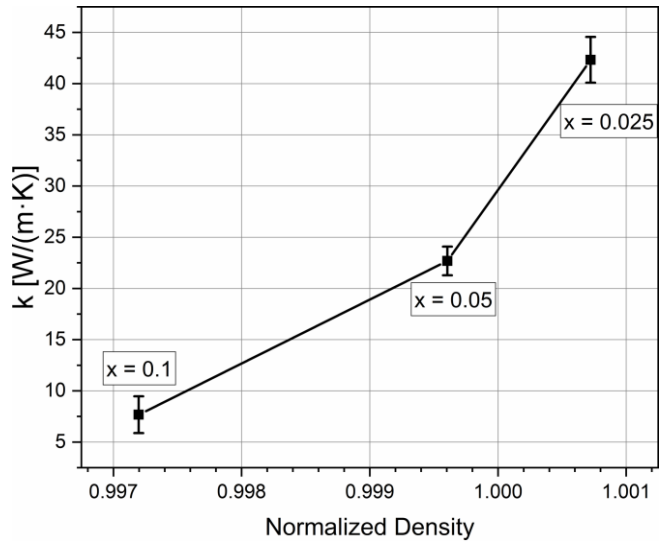

(a)

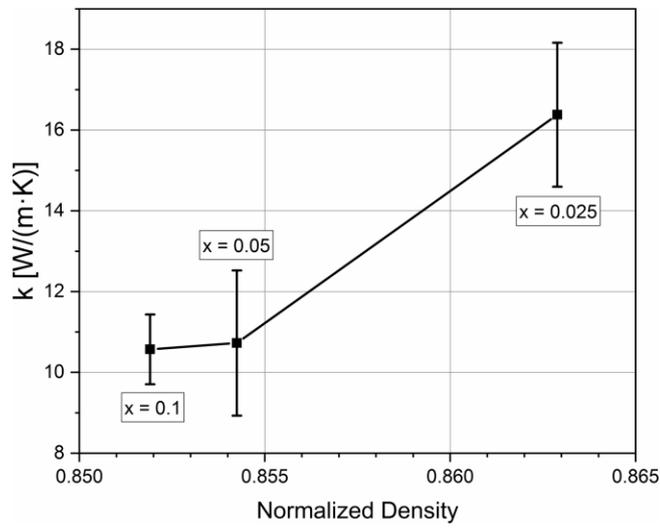

(b)

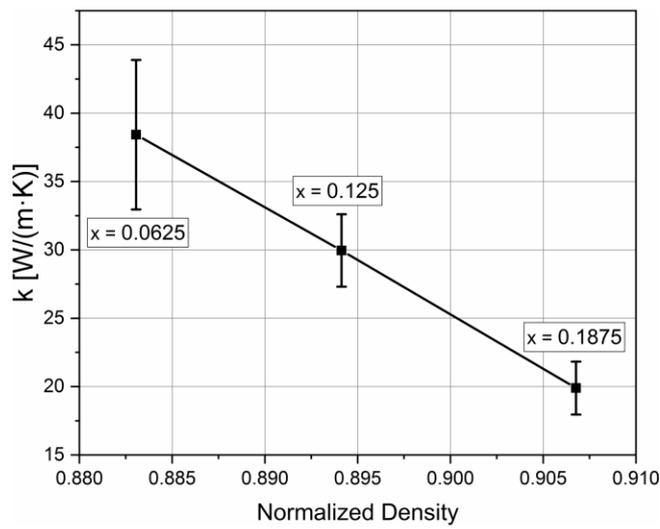

(c)

**Figure 6: Thermal conductivity as a function of normalized density, $\rho_{normalized}$, as defined by Eq. (5) for the cases of In-decorated screw dislocations in bulk GaN (a), pristine $In_xGa_{1-x}N$ random alloy (b) and pristine**

**In$_x$Ga$_{1-x}$N ordered alloy (c). The desired In concentration in the pristine order compounds was achieved by replacing Ga atoms with In ones in bulk wurtzite GaN on the energetically favorable lattice sites [80].**

The phonon Density of States (DOS) plots for the cases of bulk GaN and InN, In-decorated screw dislocations on bulk GaN and random pristine In$_x$Ga$_{1-x}$N alloy, for In content, x=0.025, 0.05 and 0.1 and ordered pristine In$_x$Ga$_{1-x}$N alloy for In content, x=0.0625, 0.125, 0.1875 and 0.25 are shown on Figure 7. GaN phonon acoustic modes are redshifted compared to their respective ones in InN by 40 – 70 cm$^{-1}$. Results are consistent with both experimental [100, 101] and theoretical [102, 103] studies of lattice dynamics for InN and GaN. The presence of In atoms in the dislocations is not so important. We observe just a redshift of the LO modes due to the presence of In and the appearance of some modes in the gap (~600-700 cm$^{-1}$). Random and order alloys have more broadened modes compared to the binary compounds, along with a small redshifting of the LO modes. For the case of ordered alloy, we observe the appearance of secondary peaks due to the ordering, an effect that characterizes it as a kind of new material.

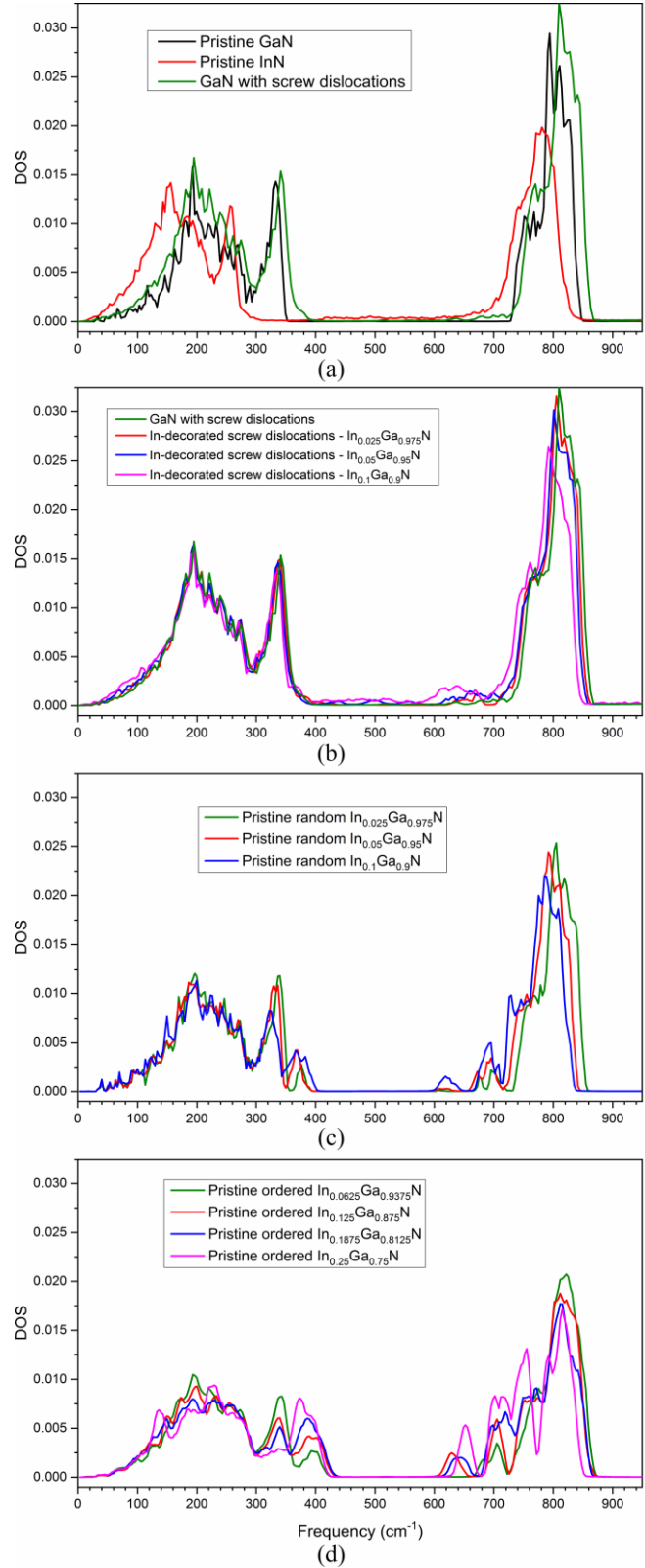

**Figure 7: Phonon density of states plot for bulk pristine GaN (black) and InN (red) and bulk GaN with screw dislocations (green) (a), In-decorated screw dislocations in bulk GaN for In content x=0 (green), 0.025 (red), 0.05 (blue) and 0.1 (magenta) (b), random pristine $In_xGa_{1-x}N$ alloy for x=0.025 (green), 0.05 (red) and 0.1 (blue) (c) and ordered pristine $In_xGa_{1-x}N$ alloy, for x=0.0625 (green), 0.125 (red), 0.1875 (blue) and 0.25 (magenta) (d).**

**Conclusions**

In this work, Equilibrium Molecular Dynamics simulations via the Green – Kubo approach and the Stillinger – Webber interatomic potential were employed for the investigation of the impact of In segregation into the cores of screw dislocations in wurtzite GaN on the thermal properties of the $In_xGa_{1-x}N$ alloy. This effect of the formation of effective nanowires in the dislocation cores is found to have a substantial impact on the material properties, as it is associated to a significant reduction of the thermal conductivity, compared to pristine GaN and InN, with the anisotropic character of k being enhanced for lower In content. Higher In concentration levels result to a further decrease of the total k almost by a factor of four, compared to GaN, along with a reduction of the anisotropy between its in-plane and c-direction components. The role of In-rich screw dislocations in the reduction of the effective k becomes evident for In compositions equal and greater than x=0.05, as for lower In content, k is similar compared to the values for the case of non-decorated screw dislocations in GaN [19]. The same In composition, moreover, for the case of In-rich screw dislocations, presents the highest anisotropy compared to the rest of the cases studied in this work, as the c axis component is almost five times higher than the in-plane one. This effect can be associated to the poor stress accommodation from the bulk due to the high In content, leading to the increase of tensile stress field components within the bulk. Relevant studies in silicon and diamond nanostructures have also verified a continuous increase of k with increasing tensile strain [97].

Regarding the case of random alloys, we observe that k appears isotropic for In composition greater than 5%. Moreover, for low In concentration in screw dislocations in GaN, k is higher than the $In_xGa_{1-x}N$ random alloy. However, for In composition higher than 10%, we observe that decorated dislocations yield sub-alloy thermal conductivity. At this point we want to discuss the impact of our results for thermoelectric applications. Alloys in general exhibit very low thermal conductivity and to overpass the alloy limit is essential when the strategy of reducing k is being followed. Furthermore, in decorated dislocations systems, the majority of the lattice remains crystalline with a small lattice perturbation around the dislocation. We could not study the electronic properties of the decorated dislocations in GaN mainly due to the extremely computationally expensive ab initio calculations for such density of dislocations, but we believe that the electronic conductivity and the Seebeck coefficient would be less severely impacted by the presence of dislocations than for alloy. Thus decorated dislocations engineering might open up new possibilities and interesting candidates for thermoelectric devices.

**Acknowledgements**

This work was supported by computational time granted from the Greek Research & Technology Network (GRNET) in the 'ARIS' National HPC infrastructure under the project NANO2D (pr006039). Ph.K. and J.K. acknowledge the support of this work from the project "INNOVATION-EL" (MIS 5002772) which is implemented under the Action "Reinforcement of the Research and Innovation Infrastructure," funded by the Operational Programme "Competitiveness, Entrepreneurship and Innovation" (NSRF 2014-2020) and co-financed by Greece and the European Union (European Regional Development Fund).